\documentclass[journal=nalefd]{achemso}
\setkeys{acs}{articletitle=true}

\DeclareUnicodeCharacter{3B1}{\ensuremath{\alpha}}

\usepackage{graphicx}
\usepackage{hyperref}
\usepackage[version=3]{mhchem}
\hypersetup{
  colorlinks,
  citecolor=blue,
  linkcolor=blue,
  urlcolor=blue}

\title{Atomic and Electronic Structure of Strongly Charged Domain Walls in van der Waals \texorpdfstring{$\alpha$-\ce{In2Se3}}{a-In2Se3}}

\author{Gillian Nolan}
\affiliation{Department of Materials Science and Engineering, The Grainger College of Engineering, University of Illinois Urbana-Champaign, Urbana, IL 61801, USA}

\author{Edmund  Han}
\affiliation{Department of Materials Science and Engineering, The Grainger College of Engineering, University of Illinois Urbana-Champaign, Urbana, IL 61801, USA}

\author{Shahriar Muhammad Nahid}
\affiliation{Department of Mechanical Science and Engineering, The Grainger College of Engineering, University of Illinois Urbana-Champaign, Urbana, IL 61801, USA}

\author{Patrick Carmichael}
\affiliation{Department of Materials Science and Engineering, Grainger College of Engineering, University of Illinois Urbana-Champaign, Urbana, IL 61801, USA}

\author{Arend M. van der Zande}
\affiliation{Department of Materials Science and Engineering, The Grainger College of Engineering, University of Illinois Urbana-Champaign, Urbana, IL 61801, USA}
\alsoaffiliation{Department of Mechanical Science and Engineering, The Grainger College of Engineering, University of Illinois Urbana-Champaign, Urbana, IL 61801, USA}
\alsoaffiliation{Materials Research Laboratory, University of Illinois Urbana-Champaign, Urbana, IL 61801, USA}

\author{Andr\'e Schleife}
\email{schleife@illinois.edu}
\affiliation{Department of Materials Science and Engineering, The Grainger College of Engineering, University of Illinois Urbana-Champaign, Urbana, IL 61801, USA}
\alsoaffiliation{Materials Research Laboratory, University of Illinois Urbana-Champaign, Urbana, IL 61801, USA}
\alsoaffiliation{National Center for Supercomputing Applications, University of Illinois Urbana-Champaign, Urbana, IL 61801, USA}

\author{Pinshane Y. Huang}
\email{pyhuang@illinois.edu}
\affiliation{Department of Materials Science and Engineering, The Grainger College of Engineering, University of Illinois Urbana-Champaign, Urbana, IL 61801, USA}
\alsoaffiliation{Materials Research Laboratory, University of Illinois Urbana-Champaign, Urbana, IL 61801, USA}

\begin{document}
\newpage

\begin{abstract}
Here, we use atomic resolution scanning transmission electron microscopy (STEM) and first principles calculations to study the atomic and electronic structure of strongly charged domain walls in $\alpha$-\ce{In2Se3}. STEM imaging and density functional theory (DFT) show that head-to-head (HH) domain walls contain a layer of nonpolar $\beta$-\ce{In2Se3}, whereas tail-to-tail (TT) domain walls are atomically abrupt.  We apply 4D STEM and multislice electron ptychography to map ferroelectric domains in 2D and 3D, showing that nearly $180^\circ$ domain walls exhibit complex, curved 3D structures that differ from ideal $180^\circ$ structures. Band structure calculations show localized conducting states within a $\sim$ 1 nm thick layer at both HH and TT domain walls, such as a midgap state at the $\beta$ layer of the HH domain wall. These properties make strongly charged domain walls in $\alpha$-\ce{In2Se3} excellent candidates for realizing 2D electron or hole gases and domain wall engineering in van der Waals ferroelectrics.
\end{abstract}


Charged domain walls (CDWs) in ferroelectrics are functional interfaces with potential applications in nonvolatile memory, logic, and neuromorphic computing. These domain walls form quasi-two-dimensional (2D) electron and hole gases that can be electrostatically written, erased, and moved, allowing atomic-scale control of 2D conductive regions  \cite{bednyakov_formation_2015,bednyakov_physics_2018,jiang_temporary_2018,wang_ferroelectric_2022}.  2D ferroelectrics such as $\alpha$-\ce{In2Se3} are of particular interest because they can maintain their polarization down to a monolayer \cite{xue_room-temperature_2018,ding_prediction_2017} and can be easily stacked and integrated into van der Waals devices \cite{huang_two-dimensional_2022,si_ferroelectric_2019}, unlocking the ability to create complex domain structures \cite{han_bend-induced_2023} and charged domain walls by design \cite{nahid_field-effect_2025}.

Broadly, domain walls (DWs) in $\alpha$-\ce{In2Se3} can be categorized according to their orientation --- out-of-plane, where the wall separates two different domains within the same $\alpha$-\ce{In2Se3} layer(s), and in-plane, which divide domains between adjacent layers. Most studies on $\alpha$-\ce{In2Se3} DWs have focused on out-of-plane (or nearly out-of-plane) oriented DWs, producing uncharged or weakly charged structures. Such DWs have been observed with piezoresponse force microscopy (PFM) \cite{zhou_out--plane_2017, xue_room-temperature_2018,xue_optoelectronic_2020,zhou_periodic_2023,yang_writing-speed_2023} and have been shown to be writable via electrical biasing \cite{xue_room-temperature_2018}, light illumination \cite{xue_optoelectronic_2020,parker_optical_2023}, and mechanical stress \cite{yang_writing-speed_2023, han_bend-induced_2023}. The structure \cite{kang_domain-wall_2021}, kinetics, and migration of these DWs have been investigated theoretically \cite{ding_prediction_2017,lu_domain_2023,bai_intrinsic_2024}. First principles calculations predict an increased band gap, resistance, and carrier localization resulting from out-of-plane DWs in $\alpha$-\ce{In2Se3} \cite{lu_domain_2023}.

In-plane DWs in $\alpha$-\ce{In2Se3} are much less studied. The layered structure of \ce{In2Se3} makes PFM studies of in-plane oriented DWs difficult, although serial layer removal has been used to access buried DWs and measure their three-dimensional (3D) profiles \cite{lu_3d_2024}.  Recently, the atomic structure and migration of DWs were studied via correlated \textit{in-situ} scanning transmission electron microscopy (STEM) and transport measurements that show distinct switching mechanisms for the two domain-wall types \cite{wu_stacking_2024}.

In-plane DWs are particularly interesting in $\alpha$-\ce{In2Se3} because they should be strongly charged, potentially creating nanometer-thick conducting layers embedded within a van der Waals semiconductor. Charged DWs form when the change in polarization across a DW has a component perpendicular to the DW, and are either head-to-head (HH) with polarizations pointing towards the DW, or tail-to-tail (TT) with polarizations pointing away from the DW. In-plane DWs in $\alpha$-\ce{In2Se3} have polarization vectors oriented $180^\circ$ to each other and orthogonal to the wall, making them strongly charged. Unlike in 3D-bonded ferroelectrics, the layered structure of $\alpha$-\ce{In2Se3} means that strongly charged DWs will be located within the basal plane of the material, potentially producing changes in the atomic and electronic structures of one or few layers that could be harnessed for novel electronic devices.
Understanding the atomic and electronic structure of these DWs --- and how they are correlated --- is critical for predicting and controlling their behavior and designing such new device concepts.

Here, we use atomic-resolution STEM imaging, 4D-STEM (four-dimensional STEM), and electron ptychography to study the 2D and 3D structure of strongly charged DWs in $\alpha$-\ce{In2Se3}.  By combining these experimental studies with density functional theory (DFT) simulations, we investigate the energetic stability and electronic properties of strongly charged DWs in $\alpha$-\ce{In2Se3}.

\begin{figure*}[tb!]
\centering
\includegraphics[width=7in]{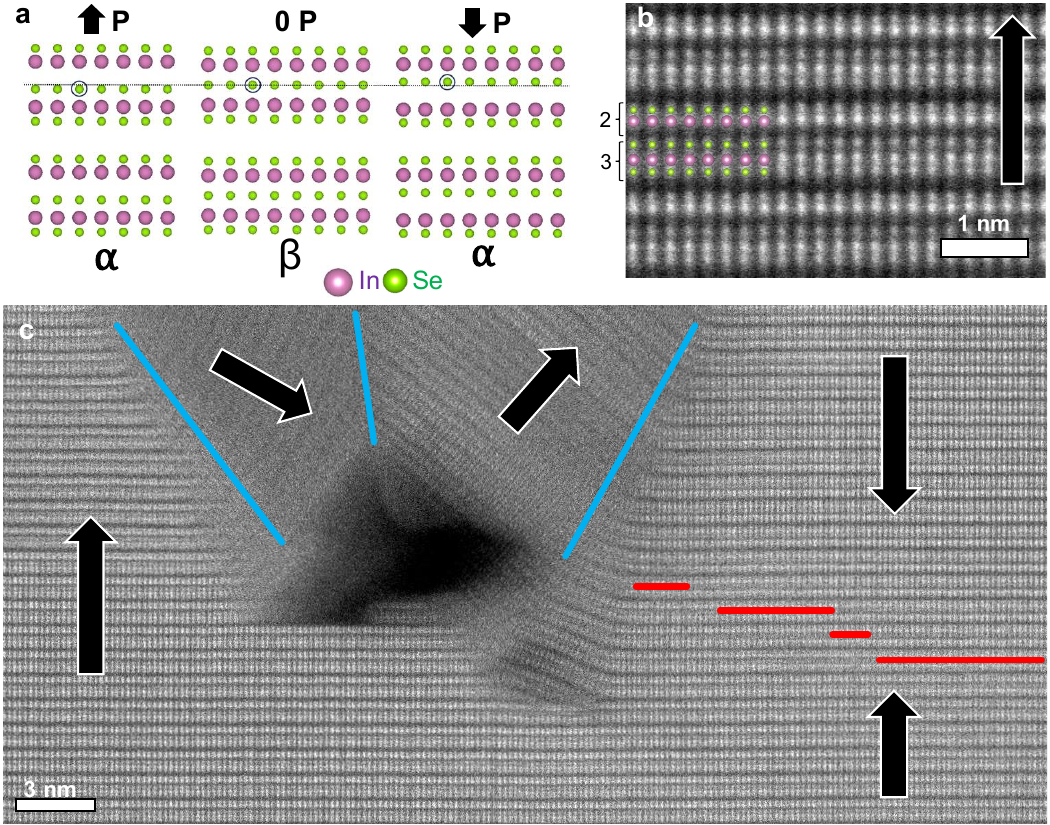}
\caption{\label{fig:In2Se3_basics}
Atomic structure and polarization of $\alpha$-\ce{In2Se3}. 
(a) Schematic of ferroelectric $\alpha$-\ce  and nonpolar $\beta$-\ce{In2Se3}, viewed along [$1\bar{1}00$]. A horizontal line through the top \ce{In2Se3} quintuple layers marks the centerline of the layer, highlighting the offset of the center Se ion in each polarization state. 
(b) Atomic resolution ADF-STEM image and corresponding polarization determined by center Se ion offset. 
(c) ADF-STEM image of ferroelectric domains around a kink, with arrows indicating polarization obtained from STEM imaging.  Transverse domain walls (blue) form where polarization within individual layers changes at kinks in the \ce{In2Se3}. The upper \ce{In2Se3} layers delaminate and undergo an odd number of kinks, creating a head-to-head domain wall (red).  }
\end{figure*}

Figure \ref{fig:In2Se3_basics} details how we fabricate and image DWs in  $\alpha$-\ce{In2Se3}. \ce{In2Se3} exhibits multiple polymorphs \cite{kupers_controlled_2018}, including ferroelectric $\alpha$-\ce{In2Se3}, which is the stable room temperature phase, and nonpolar $\beta$-\ce{In2Se3} (Figure \ref{fig:In2Se3_basics}a), which is energetically close to $\alpha$-\ce{In2Se3} and whose growth can be stabilized at room temperature via strain \cite{han_phase-controllable_2023}. 
In both phases, the atoms are arranged in quintuple layers (hereby referred to as layers) composed of Se-In-Se-In-Se.  The $\alpha$-\ce{In2Se3} phase is further divided into two stable configurations, distinguished by the stacking alignment between quintuple layers (see SI Figure S1): 2H (space group $R3m$) comprising 2 antiparallel layers, and 3R, comprising 3 parallel layers (space group $P6_3mc$).  

As illustrated in Figure \ref{fig:In2Se3_basics}a, out-of-plane polarization can be determined by tracking shifts of the middle Se plane: in the nonpolar $\beta$ phase, the middle Se is centered within each \ce{In2Se3} layer.  In the ferroelectric $\alpha$ phase, the middle Se plane is displaced by 36 pm along the c-axis. This shift can be readily measured in atomic-resolution STEM, providing a direct measure of polarization direction (Figure \ref{fig:In2Se3_basics}b).  While some works discuss an in-plane component to the electric polarization in $\alpha$-\ce{In2Se3}, the material exhibits three-fold in-plane symmetry, which should preclude in-plane ferroelectricity \cite{xiao_intrinsic_2018,bai_intrinsic_2024} in the absence of strain. We therefore refer to the total polarization and out-of-plane polarization interchangeably in this work.

To investigate the atomic structure of ferroelectric DWs in $\alpha$-\ce{In2Se3}, we fabricated samples as follows. Bulk $\alpha$-\ce{In2Se3} crystals were mechanically exfoliated onto Si/SiOx substrates, producing 10-100 nm thick flakes of $\alpha$-\ce{In2Se3}. We previously showed that mechanical stress from this process embeds kinks and out-of-plane DWs within the $\alpha$-\ce{In2Se3} \cite{han_bend-induced_2023}. Next, we fabricated TEM (transmission electron microscopy) cross-sections using focused ion beam (Methods in SI section 1.1). Figure \ref{fig:In2Se3_basics}c shows an annular dark-field (ADF)-STEM image of the resulting structure (Methods in SI section 1.2).  In this region, the top layers of the sample undergo a kink, leaving an $\sim$ 6 nm void in the sample, while the bottom layers remain undisturbed. 

By determining the local polarization from the Se plane shifts, we observe two distinct domain wall types. Out-of-plane, or transverse, DWs form where polarization changes within layers at kinks in the \ce{In2Se3}, as previously observed by Han et al.\cite{han_bend-induced_2023}. The bound charge at ferroelectric interfaces is given by the change in the normal component of the polarization across the domain wall\cite{bednyakov_physics_2018}; correspondingly, the transverse DWs observed in Figure \ref{fig:In2Se3_basics} are uncharged or weakly charged.

We also observe strongly charged DWs: 180$^\circ$, or in-plane DWs along the stacking direction.  In Figure \ref{fig:In2Se3_basics}c, the upper, delaminated \ce{In2Se3} layers undergo an odd number of kinks, leading to a net change in polarization from up to down across the image. 
Meanwhile, the bottom layers maintain their polarization. This produces a $180^\circ$ domain wall where the top and bottom domains meet on the right side of the image, with polarizations pointing inwards in a HH orientation. We observe both HH DWs and TT DWs near kinks in \ce{In2Se3}.  TT DWs are less common: in 8 samples, we found three unique TT and seven unique HH CDWs.

\begin{figure*}[tb!]
\centering
\includegraphics[width=6in]{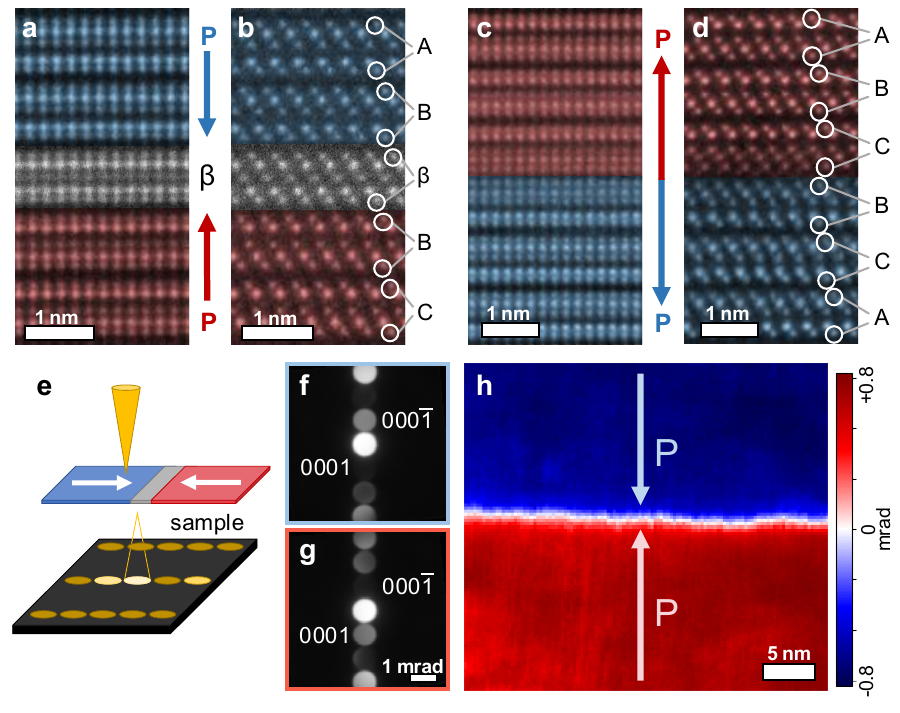}
\caption{\label{fig:wall_structure}
Atomic structure and polarization of strongly charged domain walls in $\alpha$-\ce{In2Se3}.  
(a,b) ADF-STEM images of head-to-head DWs imaged along (a) $[1\bar{1}00]$ and (b) $[10\bar{1}0]$, with red and blue shading indicating local polarization direction. The head-to-head domain wall contains one layer of nonpolar $\beta$-\ce{In2Se3}. 
(c,d) ADF-STEM images of tail-to-tail DWs imaged along (c) $[1\bar{1}00]$ and (d) $[10\bar{1}0]$. No nonpolar layer is visible in the TT DW.  
A/B/C lettering in (b,d) indicates stacking order. The HH and TT domain walls are each accompanied by in-plane shifts between quintuple layers, which preserve the relative local alignment between outer Se planes across the van der Waals gaps.
(e) Schematic of 4D-STEM acquisition. A 2D array of diffraction patterns is collected as a function of position on the sample.
(f) A representative PACBED pattern from the downwards polarized region of (h) with $(0001)$ and $(000\bar{1})$ Bragg disks marked, where intensity is greater in the $(000\bar{1})$ disk.
(g) A representative PACBED pattern from the upwards polarized region of (h). Intensity is greater in the $(000\bar{1})$ disk. 1 mrad convergence angle at 300 kV is used for (f,g).
(h) 4D-STEM polarization map from center of mass of intensity in (0001) disks. The domain wall is clearly visible in the CoM image, providing a rapid, large-area method to locate domain walls in \ce{In2Se3}.
}
\end{figure*}

Figures \ref{fig:wall_structure}a-d show ADF-STEM images comparing the structures of HH and TT CDWs. Figure \ref{fig:wall_structure}a shows a HH DW, with ferroelectric domains separated by a single layer of the nonpolar $\beta$ phase. We find this $\beta$ layer in every HH DW observed. Depolarized layers such as this are common at strongly charged DWs because the abrupt change in polarization produces a strong electric field, which depolarizes the material around it. Unlike in conventional ferroelectrics, however, where depolarization typically occurs over several nanometers \cite{catalan_domain_2012,bednyakov_physics_2018}, we observe that $\alpha-\ce{In2Se3}$ forms a highly localized (1 nm) nonpolar layer localized to a single layer of \ce{In2Se3}.

In contrast, the TT CDW in Figure \ref{fig:wall_structure}c appears atomically abrupt. Unlike our HH structures, our STEM images show no sign of a nonpolar $\beta$ phase.  Instead, we observe decreased contrast in the \ce{In2Se3} layers on either side of the interface, likely arising from local strain or defects. This abrupt structure deviates from conventional ferroelectric DWs:  bulk and single crystal $\alpha$-\ce{In2Se3} exhibit n-type conductivity \cite{nahid_field-effect_2025,kremer_quantum_2023}, and  n-type semiconductors typically have wider TT CDWs than HH walls due to the lower carrier concentration of screening charges \cite{catalan_domain_2012}.

In Figures \ref{fig:wall_structure}b and \ref{fig:wall_structure}d, we investigate the interlayer stacking arrangements at DWs. Figures \ref{fig:wall_structure}b and \ref{fig:wall_structure}d show HH and TT DWs viewed along the $[10\bar{1}0]$ zone axis.  Away from the DWs, our samples are in one of two uniform stacking sequences: 3R (ABC stacking) or 2H (AB' stacking).  In contrast, the HH CDW in Figure \ref{fig:wall_structure}b has a stacking order of AB$\beta$BC (see SI Figure S2 for broader field of view). Assuming no stacking faults were initially present, forming this structure would require sliding one domain by 1/3 unit cell, from the A to B stacking configuration.  This shift preserves the relative local alignment between outer Se planes across the van der Waals gaps, as indicated by the circled atomic columns in Figure \ref{fig:wall_structure}b. We observe this interlayer shift consistently across all HH CDWs (SI Figure S3).  We also observe stacking shifts at TT DWs, shown in Figure \ref{fig:wall_structure}d for the 3R stacking and in SI Figure S4 for 2H.  Because the domains in our samples are formed under mechanical stress during exfoliation, their formation also provides a natural pathway for interlayer shear and sliding.  As we show in DFT calculations later on, these stacking shifts are energetically favorable for the HH 3R structure. 

Our experimental results differ significantly from recently reported CDW structures in $\alpha$-\ce{In2Se3}, where HH and TT CDWs were suggested to occur only in 2H $\alpha$-\ce{In2Se3} \cite{wu_stacking_2024} and not 3R $\alpha$-\ce{In2Se3}. These claims were based on an argument that interlayer slip does not occur. In contrast, we observe HH and TT CDWs in both 3R and 2H samples (SI Figures 3,4) - in fact, of CDWs imaged along the $[10\bar{1}0]$ direction where the 3R vs 2H can be distinguished, five out of six were found in 3R $\alpha$-\ce{In2Se3}.

Next, we used 4D-STEM center-of-mass (CoM) imaging to map DWs in \ce{In2Se3} over larger, nanoscale regions. In 4D-STEM, an array of convergent beam electron diffraction (CBED) patterns are collected (Figure \ref{fig:wall_structure}e).  Figures \ref{fig:wall_structure}f and \ref{fig:wall_structure}g show position-averaged CBED (PACBED) patterns collected on each side of a HH domain wall. The broken inversion symmetry in the atomic structure of $\alpha$-\ce{In2Se3} \cite{tanaka_electron_1964} produces a corresponding broken symmetry in the intensity of the $(0001)$ and $(000\bar{1})$ Bragg disks.
We use center of mass (CoM) imaging to track polarization based on these intensity shifts. We first apply virtual apertures to isolate the $(0001)$ and $(000\bar{1})$ diffraction disks, then compute the CoM of the two disks at each scan position. Figure \ref{fig:wall_structure}h plots the CoM shifts along the $[000l]$ direction for each CBED pattern at a HH CDW. The polarization change at the domain wall is clearly visible in the CoM image. These results highlight how 4D-STEM can be used to rapidly locate ferroelectric DWs in \ce{In2Se3} at larger length scales and without requiring atomic resolution imaging.

\begin{figure*}[tb!]
    \centering
    \includegraphics[width=3.25in]{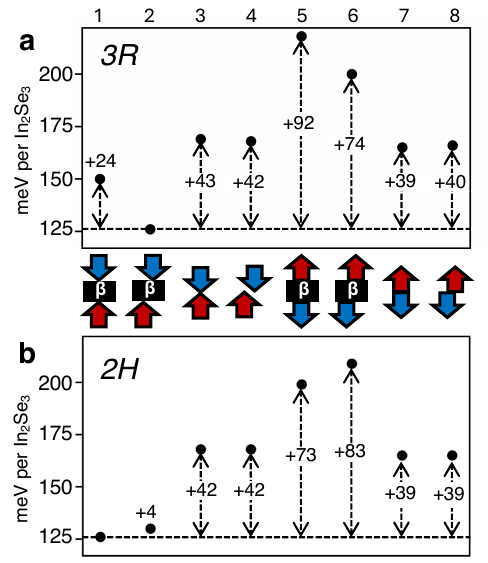}
    \caption{Relative domain wall energies of head-to-head and tail-to-tail DW structures in 
    (a) 3R and 
    (b) 2H $\alpha$-\ce{In2Se3}, 
    both with and without a nonpolar $\beta$ layer. 
    From left to right, the structures are: 1) HH with $\beta$ and no stacking shift (SS), 2) HH with $\beta$ and SS, 3) abrupt HH, 4) abrupt HH with SS, 5) TT with $\beta$, 6) TT with $\beta$ and SS, 7) abrupt TT, and 8) abrupt TT with SS. The lowest energy HH (but not TT) domain wall structures contain a $\beta$ layer. 
    Both HH and TT DWs in 3R with $\beta$ are lower energy with a stacking shift, whereas DWs with and without stacking faults are similar in energy for the other DW configurations. These results are consistent with the structures  observed in our experiments. }
    \label{fig:energetics}
\end{figure*}

\begin{figure*}[tb!]
\centering
\includegraphics[width=6in]{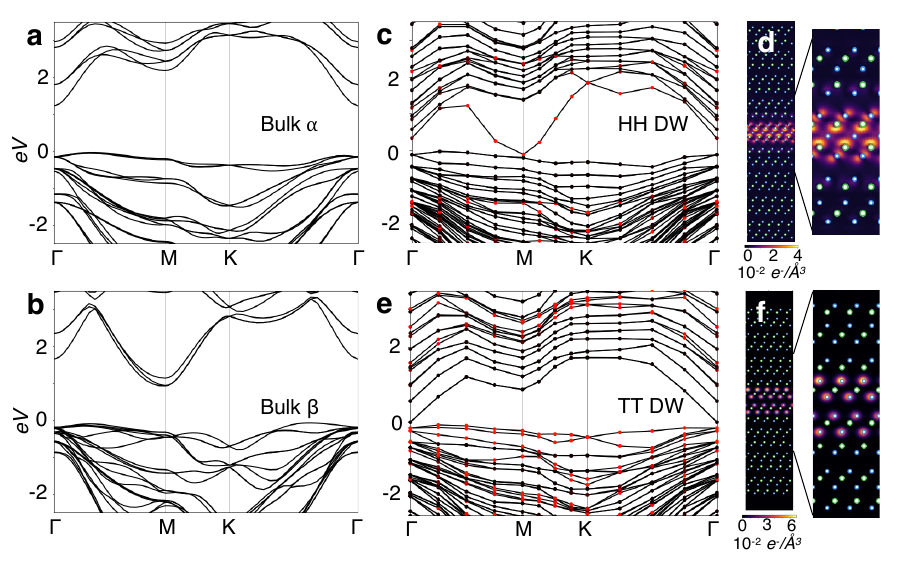}
\caption{\label{fig:dft}
Simulated electronic structure in bulk and interfaces in \ce{In2Se3} from Density Functional Theory (hybrid, HSE06 functional). 
Electronic band structures for (a) bulk $\alpha$-\ce{In2Se3} with a 1.26 eV indirect band gap and 
(b) bulk $\beta$-\ce{In2Se3} with a 0.91 eV indirect gap. 
(c) Electronic band structure of a 9 layer slab with head-to-head domain wall in $\alpha$-\ce{In2Se3}. States colored red denote spatial localization within the $\beta$ layer of the head-to-head domain wall, obtained via partial charge density calculations. 
(d) Partial charge density map showing the gap-crossing HH-DW state at the M point (i.e. the conduction band minimum in (c)) with atomic positions superimposed. 
(e) Electronic band structure of an 8 layer slab with tail-to-tail domain wall. States colored red denote localization within the two innermost layers of the tail-to-tail domain wall. 
(f) Partial charge density map showing localization of the valence band maximum in (e) with atomic positions superimposed. 
These simulations show that the domain walls strongly modify the local electronic structure in ferroelectric \ce{In2Se3}, producing mid-gap states localized within 1 nm of the DW interface.
}
\end{figure*}

Figures \ref{fig:energetics} and \ref{fig:dft} explore the energetic stability and electronic properties of charged DWs in $\alpha$-\ce{In2Se3} from Density Functional Theory (DFT). We relaxed a series of slab structures containing HH and TT DWs, each containing equal numbers (3-6) of upwards and downwards polarized 3R or 2H $\alpha$-\ce{In2Se3}, with either one or no $\beta$ layer in the middle.  As shown in Figure \ref{fig:energetics}a, we simulated four HH DW types for 3R stacking. The lowest-energy structure contained both a $\beta$ layer and stacking shift, matching the structure observed in experiment. Omitting the stacking shift raises the energy by 24 meV per in-plane formula unit, and removing the $\beta$ layer increases the energy by $\sim$ 40 meV/formula unit regardless of stacking. For the TT DW in 3R, the $\beta$-free interface is most energetically favorable, again matching the observed experimental structures  (details in SI section 1.3-4). 

2H simulations exhibit similar trends --- adding a $\beta$ layer lowers the energy of the HH structure, but not the TT. However, adding slip to the 2H structures do not significantly alter their energetic stability (Figure \ref{fig:energetics}b).  We also note that TT DWs are less stable than the HH DWs by roughly 40 meV/formula unit.  This is consistent with our experimental observation that HH DWs are more common.

Figure \ref{fig:dft} shows electronic structure calculations from DFT. Figures \ref{fig:dft}a and \ref{fig:dft}b show the band structures of bulk $\alpha$-\ce{In2Se3} and $\beta$-\ce{In2Se3} calculated using an HSE06 hybrid functional \cite{heyd_hybrid_2003,heyd_erratum_2006}. For bulk $\alpha$-\ce{In2Se3}, we obtain a 1.37 eV direct band gap (1.26 indirect), close to the experimental bulk direct gap of 1.39 eV \cite{si_ferroelectric_2019} (see SI Figure S5 for bulk band structures). 

Figure \ref{fig:dft}c shows the band structure for the lowest-energy HH domain wall --- one with both a $\beta$ layer and stacking shift --- calculated using an HSE06 hybrid functional. Compared to the bulk $\alpha$ phase, we observe a new state that crosses the band gap. Partial charge density calculations in (Fig \ref{fig:dft}c-d) reveal that this HH-DW state is localized to the $\beta$ layer (see SI Figure S6 for local densities of states for each layer). 

Figure \ref{fig:dft}e shows the band structure for the lowest-energy TT domain wall.  This structure does not contain a $\beta$ layer or corresponding midgap state. However, band bending due to the potential gradient reduces the gap significantly, to 0.07 eV. By integrating the partial charge density over the two innermost layers of $\alpha$-\ce{In2Se3}, we find that the states at the top of the valence band are strongly localized to the domain wall. 
Figure \ref{fig:dft}f plots the partial charge density at the valence band maximum, showing a high density of states confined within $\sim$ 1 nm of the domain wall. PBE-computed local DOS for the TT wall are in SI Figure S7. Together, our simulations indicate that the electronic structure of $\alpha$-\ce{In2Se3} is strongly modified by the presence of CDWs. Both the  HH and TT CDW structures produce localized conducting states within a highly localized, 1 nm-wide region at the domain wall (further details in SI section 1.6). 

\begin{figure*}[bt!]
\centering
\includegraphics[width=6.25in]{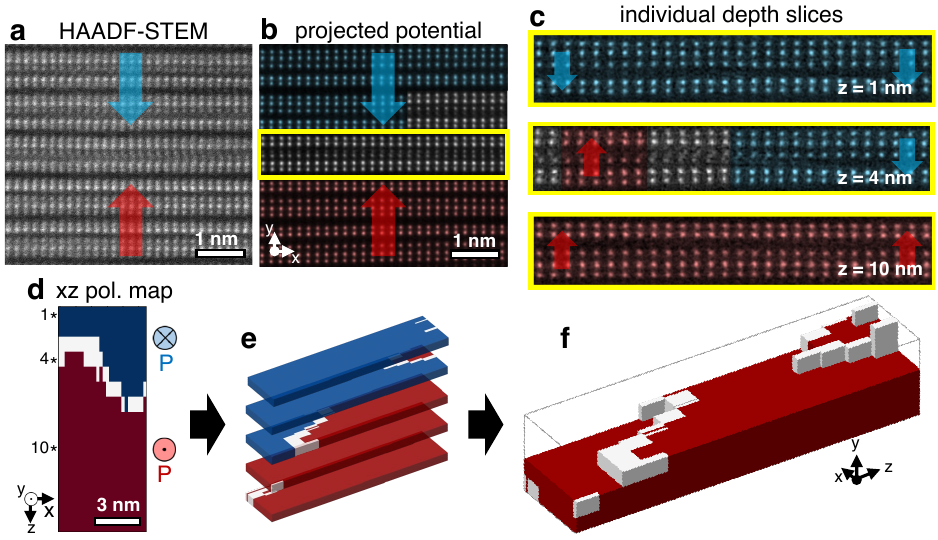}
\caption{\label{fig:ptycho}
Multislice electron ptychography reconstruction of a nearly $180^\circ$ \ce{In2Se3} domain wall with in-plane and out-of-plane components. 
(a) ADF-STEM image of a 3D domain wall, where polarization in some layers is unclear in projection. 
(b) 2D projection of ptychographic reconstruction of a nearby region to (a). Six atomic planes appear in the highlighted layer, indicating both up and down polarizations stacked in projection. Unlike perfectly oriented HH domain walls, no $\beta$ layer is evident.
(c) Depth slices of the highlighted layer in (a), taken at depths of z=1 nm, 4 nm, and 10 nm. The polarization is uniformly downwards at z = 1 nm and upwards at z = 10 nm.  The z = 4 nm image shows the polarization changing from up (red) to down (blue) across the image, with a central transitional region $\sim$1 nm thick and another transition beginning at the far left. 
(d) Polarization map of the layer in (b), orientated in the xz plane. Blue/red indicate polarization into/out of the xz plane and white represents regions where the polarization cannot be assigned. The domain wall migrates between layers and exhibits a complex, curved structure within layers. Pixel size in x corresponds to interatomic spacing (2.1 \AA) and pixel size in z corresponding to reconstruction slice thickness (1 nm). The 1, 4, and 10 nm slice depths in (c) are indicated.
(e) Stack of xz polarization maps for all 5 layers in (a).
(f) 3D volume displaying polarization, with transparent upper domain. Although the polarization is oriented near a head-to-head configuration, the atomic structure domain wall structure is complex and curved in 3D and distinct from the HH structure. 
}
\end{figure*}

Next, we investigate a complex, 3D-structured domain wall, as shown in the ADF-STEM image in Figure \ref{fig:ptycho}a.  We reconstructed its 3D structure using multislice electron ptychography (MEP) with 2-3 nm depth resolution \cite{chen_electron_2021,maiden_ptychographic_2012} (details in SI section 1.7). Figure \ref{fig:ptycho}b shows a 2D projection of the  MEP reconstruction. The structure contains a nearly $180^\circ$ HH domain wall. Unlike ideal $180^\circ$ HH DWs, this tilted domain wall does not contain $\beta$-\ce{In2Se3}. Instead, 6 atomic planes appear in the highlighted layer instead of 5, indicating both up and down polarizations stacked in projection. These central Se sites are separated by 76 pm, difficult to resolve ADF-STEM but easily distinguishable with the increased resolution of electron ptychography. 

Figure \ref{fig:ptycho}c shows individual depth slices at z = 1, 4, and 10 nm. At z = 1 nm, the polarization points downwards (blue) across the full field of view. At 4 nm, polarization towards the left half of the image has begun to switch upwards (red), with an $\sim$ 1 nm transition region between up and down polarization where the middle Se atomic position and polarization are unclear and another transition region at the far left. By z = 10 nm, the polarization has fully switched upwards.

Figure \ref{fig:ptycho}d shows a corresponding top-down polarization map of the highlighted layer in \ref{fig:ptycho}b/c, showing the two domains coexisting within this layer with depths at z = 1, 4, and 10 nm indicated. To generate this map, we determined the polarization of each pixel by tracking the Se displacement in each unit cell through each slice using atomap\cite{nord_atomap_2017}. 
Zero polarization (white in Figure \ref{fig:ptycho}d-f) is assigned to the transitional regions.  Figure \ref{fig:ptycho}e-f shows polarization maps for all of the layer in \ref{fig:ptycho}b, with either blue (\ref{fig:ptycho}e) or transparent shading (\ref{fig:ptycho}f) indicating the up-polarized domain.  In Figure \ref{fig:ptycho}e-f, the domain wall migrates between layers as shown by the steps of interface between the red ($P_{up}$, along y) and blue/transparent ($P_{down}$) domains. Notably, the DW is not flat, but instead curves within layers and between them. Our results show the complex structure of CDWs in \ce{In2Se3}, previously viewed only at length scales of 10s of nm to microns\cite{lu_3d_2024}, persist down to the atomic scale.

The lack of $\beta$ layer and curvature of the interface in nearly HH DWs are important considerations. Transport properties in artificial HH DWs in $\alpha$-\ce{In2Se3} vary with the local atomic order at the DW, where structural changes are linked to trap states \cite{nahid_field-effect_2025}.
Our results indicate that the edges of DWs exhibit polarization texture at the nanoscale, an effect that should also impact their electronic properties.

To summarize, our work illustrates the intricate, strong coupling between atomic structure, interlayer sliding, polarization, and electronic structure in strongly charged DWs in ferroelectric $\alpha$-\ce{In2Se3}. This indicates that charged DWs in $\alpha$-\ce{In2Se3} are promising structures for hosting 2D electron and hole gases within a nm-thick layer of semiconducting van der Waals ferroelectric, a result that may lead to the realization of novel physical properties or new device concepts in these systems.

\section{\label{sec:acknowledge}Acknowledgments}
This work was funded by the National Science Foundation - Materials Research Science and Engineering Center (NSF-MRSEC) with award numbers DMR-1720633 and DMR-2309037, and the Air Force under grant number FA9550-20-1-0302. We also acknowledge funding from the Office of Naval Research under grant number ONR N00014-18-1-2605.
This work was carried out in part in the Materials Research Laboratory Central Facilities at the University of Illinois at Urbana-Champaign, and we acknowledge use of microscopy facilities funded by NSF-MRSECs under award numbers DMR-1720633 and DMR-2309037. 
This work made use of the Illinois Campus Cluster, a computing resource that is operated by the Illinois Campus Cluster Program (ICCP) in conjunction with the National Center for Supercomputing Applications (NCSA) and which is supported by funds from the University of Illinois at Urbana-Champaign. 
This work used the Delta advanced computing and data resource at the NCSA at University of Illinois Urbana-Champaign through allocation MAT240031 from the Advanced Cyberinfrastructure Coordination Ecosystem: Services \& Support (ACCESS) program \cite{boerner_access_2023}, which is supported by National Science Foundation grants \#2138259, \#2138286, \#2138307, \#2137603, and \#2138296.

\bibliography{lib_20260126.bib}
\newpage
\subsection{Graphical TOC Entry}
\begin{figure}[H]
\includegraphics[width=1.8in]{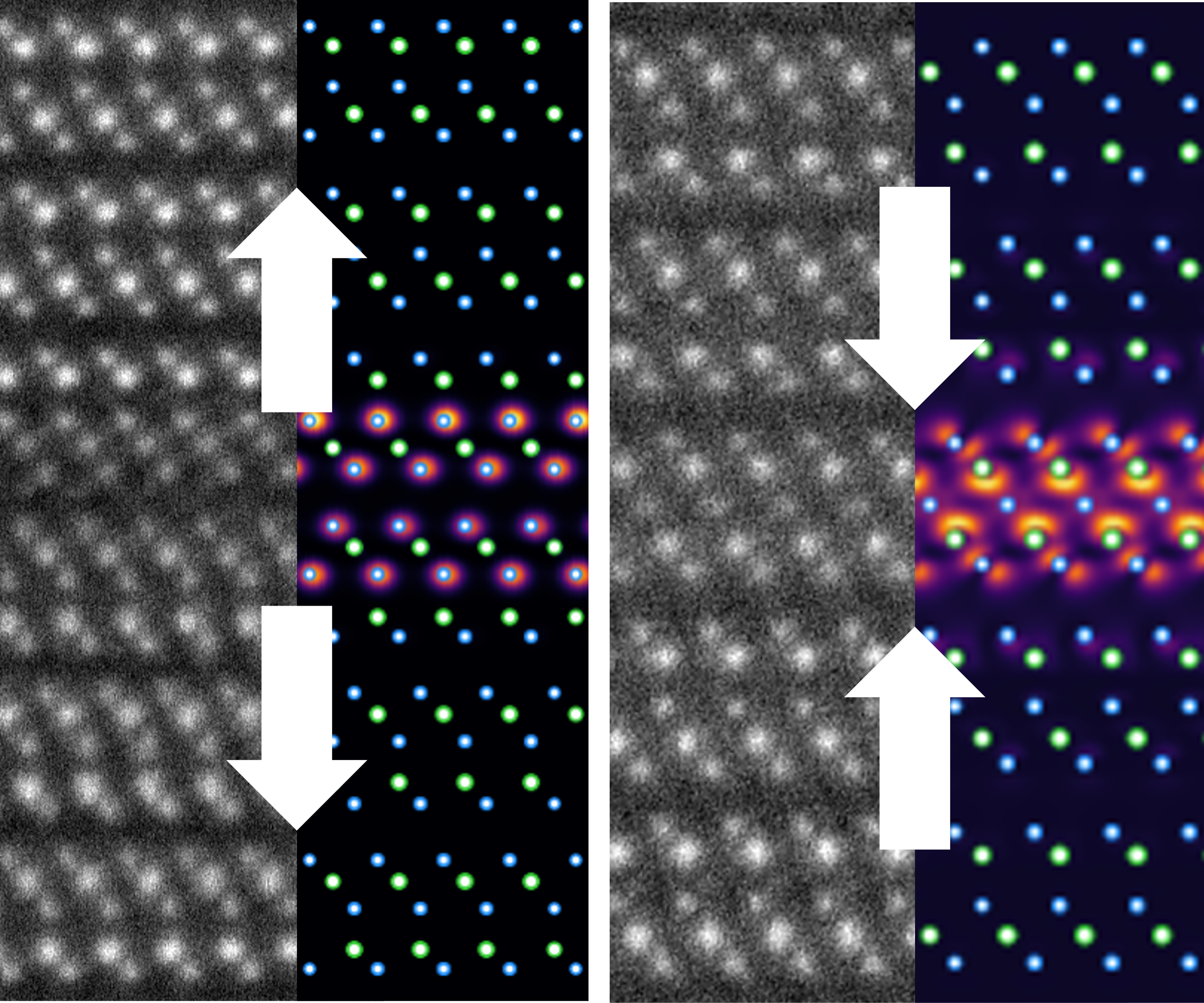}
\label{TOC}
\end{figure}

\end{document}